\documentstyle[prl,aps,epsf]{revtex}
\begin{document}
\draft
\title{ Time Reversal and Parity Breaking by Propagation }

\author{H.J\"unemann and A.Weiguny}
\address{
Institut f\"ur Theoretische Physik, Universt\"at M\"unster, 48149 M\"unster,
Germany}
\author{B.G.Giraud}
\address
{Service de Physique Th$\acute{\rm{e}}$orique, DSM--CEA Saclay, 91191
Gif/Yvette, France}
\date{\today}
\maketitle
\begin{abstract}
The time independent mean field method (TIMF) for the calculation of
matrix elements of non relativistic propagators is based on a variational
principle whose non linearity induces a multiplicity of variational
solutions. Several of them can break any symmetry shared by the Hamiltonian
and initial and final states. We describe a soluble model where, in
particular, time reversal and parity breakings occur. Such breakings
account for important properties of propagation amplitudes.
\end{abstract}

\narrowtext
The observation of symmetry breakings in solutions
of variational principles is familiar for static problems, such as those
described by the Hartree-Fock method for deformed nuclei. Similar
breakings can be expected for other variational theories, like the time
independent mean field (TIMF) theory of reactions and decay \cite{1}.
The TIMF method has already been implemented for various nuclear and atomic
processes involving 3, 4, 5 and 8 particles \cite{2} and symmetry breakings
were indeed found \cite{3}. In this letter we describe a soluble
model which completely interpolates between mean field trial states and
exact solutions. It provides a detailed observation of symmetry breakings as
well as conservation. Because propagation amplitudes are obtained by saddle
points rather than absolute maxima or minima, the multiplicity of solutions
raises an ambiguity. The model shows that actually all mean field solutions
should later be mixed linearly to recover, for the complete energy range, the
best possible approximation of the exact solution.

We start from a Schwinger-like functional \cite{1,4}
\begin{equation}
F(\Psi',\Psi) = \frac{<\chi'|\Psi> <\Psi'|\chi>}{<\Psi'|(E - H)|\Psi>},
\end{equation}
to calculate the amplitude $D(E)\equiv <\chi'|(E - H)^{-1}|\chi> $ for
propagation between initial and final states
$\chi, \chi'$. The variation of $F$ gives, with appropriate normalizations,
\begin{mathletters} \begin{eqnarray}
(E - H)|\Psi> &=& |\ \chi>, \\
<\Psi'|(E - H) &=& <\chi'|, \\
D(E) = <\chi'|\Psi> &=& <\Psi'|\chi> \, ,
\end{eqnarray} \end{mathletters} \noindent
with $(\Psi,\Psi')$ as a saddle point of $F.$ For our model
we consider a single-particle basis made of two square integrable orbitals
$|a>$ and $|b>,$ with single-particle energies $-1$ and $1,$ respectively.
In the resulting space for two distinct particles, spanned by $|a^2>,$ $|ab>,$
$|ba>,$ $|b^2>,$ we use for $H$ simply the corresponding
sum of one-body Hamiltonians, with eigenvalues $-2,$ $0,$ $0$ and $2,$
respectively. This is clearly a Hamiltonian invariant under
particle exchange (denoted parity in the following)
and complex conjugation (time reversal).
Then we choose $|\chi'>=|\chi>,$ and $\chi$ as a two-body state,
described by a product of two identical, single-particle mixtures,
\begin{equation}
|\chi>=(r|a>+|b>)_1(r|a>+|b>)_2,
%r^2|a^2>+r|a_1b_2>+r|b_1a_2>+|b^2>,
\end{equation}
where $r$ is a frozen, {\it real} parameter. Both $\chi,\chi'$ are
invariant under parity. The same holds for $\Psi,\Psi'.$ Furthermore,
since $|\chi>=|\chi'>$ and $r$ is real, time reversal symmetry is expressed
by the property $|\Psi'>=|\Psi>$ if $E$ is real, and $|\Psi'>=|\Psi^{*}>$ if
$E$ is complex, see Eqs.(2a,b).

It is seen trivially that the exact value of $D(E)$ is
\begin{equation}
D_{ex}(E)=\frac{r^4}{E+2}+\frac{2r^2}{E}+\frac{1}{E-2}.
\end{equation}
This pole structure allows $E$ to remain real in the following.
Our model investigates reconstructions of $D(E)$ with trial
functions $\Psi,$ $\Psi'$ restricted to less flexible forms,
\begin{eqnarray}
|\Phi> &=& c|\ a^2>+\ s\ |\ a_1b_2>+\ t\ |\ b_1a_2>+\ st\ |\ b^2>, \nonumber \\
<\Phi'| &=& c<a^2|+u<a_1b_2|+v<b_1a_2|+uv<b^2|,
\end{eqnarray} \noindent
where $c$ is a frozen, {\it real} parameter which expresses correlation
constraints, and $s,$ $t,$ $u,$ and $v$ are four variational parameters,
possibly complex. It is easily verified that, for the special
value $c_{ex}=E^2/(E^2-4),$ the space of trial functions $\Phi,$ $\Phi'$
contains the exact solutions $\Psi,$ $\Psi'$ given by Eqs.(2),
with $s=t=u=v=E/[r(E-2)].$ Conversely, for $c=1,$ we
generate a TIMF limit where the trial states $\Phi,$ $\Phi'$ are
uncorrelated, namely simple products of single-particle orbitals. The
model thus achieves, as a function of $c,$ an interpolation between a mean
field approximation and an exact theory. Note that, despite its simplicity,
the model is non trivial since the inverse of an additive one-body operator is
a complicated sum of many-body operators. This letter studies whether
time reversal, namely $s=u$ and $t=v,$ on one hand, and on the other hand
parity, namely $s=t$ and $u=v,$ are properties conserved or broken by the
variational principle.

With the restricted $\Phi,\Phi'$ of Eqs.(5), $F$ becomes
\begin{equation}
F=\frac{(cr^2+rs+rt+st)(cr^2+ru+rv+uv)}{E(c^2+su+tv+stuv)-2(stuv-c^2)},
\end{equation}
whose derivatives with respect to $s,t,u,v$ are easy to obtain.
Let $F_s,$ $F_t,$ $F_u,$ $F_v$ be the multivariate polynomial numerators
of these derivatives. From the conditions $F_t=F_v=0$ a straightforward,
linear elimination of $t$ and $v$ in terms of rational functions of $s$ and $u$
is possible. The substitution of such rational fractions for $t$ and $v$ inside
$F_s$ and $F_u$ then leaves two residual polynomial conditions for $s$ and $u.$
It is finally easy, albeit slightly tedious, to eliminate for instance $u$
between these two polynomials and obtain a polynomial resultant $R(s;E,r,c),$
the roots of which designate the saddle points of $F$, Eq.(6).

It turns out that, except for spurious factors which either identically cancel
$F$ or make $\Phi,\Phi'$ divergent, $R$ reduces into the product of three
simpler resolvents,
\widetext \begin{mathletters} \begin{eqnarray}
R_{cc} &=& c^2r(E+2) + c(2c+cE-Er^2)s + (E+2cr^2-cEr^2)s^3 + r(2-E)s^4, \\
R_{pb} &=& cE(2c+cE+Er^2) + r(4c^2+E^2-c^2 E^2)s + E(E-2cr^2+cEr^2)s^2, \\
R_{tb} &=& cr(2c+cE+Er^2)+c(2r^4-Er^4+E+2)s + r(2cr^2-cEr^2-E)s^2.
\end{eqnarray} \end{mathletters} \narrowtext \noindent
The first factor, $R_{cc},$ generates four saddle points for which $s=t=u=v,$
namely both parity and time reversal are conserved. The second factor,
$R_{pb},$ induces two saddle points for which $s=u$ and $t=v,$ but $s \ne t,$
hence parity is broken, while time reversal is conserved. The corresponding
value $F_{pb}$ of $F$ is the same at both such saddle points,
\begin{equation}
%F_{pb}=
\frac{[r^2(c^2r^2+c^2+2c-1)+c^2]E^2 + 2c^2(1-r^4)E - 4c^2r^2}
{E(c^2E^2-E^2-4c^2)}.
\end{equation}
The last factor, $R_{tb},$ induces two saddle points for which $s=t$ and
$u=v,$ but $s \ne u,$ hence parity conservation but time reversal breaking,
with again equal values of $F$ at both saddle points,
\begin{equation}
F_{tb}=(c-1)\frac{(cr^4+2r^2+c)E+2c(1-r^4)}{(c^2-1)E^2-4c^2}.
\end{equation}
While many combinations of the real parameters $E,r,c$ induce real values for
$s,t,u,v$ at such symmetry breaking saddle points, see Figs.(1,2) for a
{\it pb}-- and a {\it tb} saddle point, respectively, there are equally
frequent situations where those pairs $(s,t)$ or $(s,u)$ which are not
degenerate actually take complex conjugate values. Nevertheless, as shown
by both Eqs.(8,9), the functional remains real valued. This allows useful
analytic continuations of variational estimates of $D,$ which will indeed be
plotted in several forthcoming Figures, regardless of the real or complex
nature of the corresponding saddle points. Conversely, for the fully symmetric
cases described by $R_{cc,}$ real estimates of $D$ are obtained when $s$ is
real only, unless $ReF$ is used as an estimate of $D$ when $F$ is complex.

By means of Fig.3 we now discuss the behaviors of the six estimates of $D$
provided by $R$ when $c$ takes on all values between the limit without
correlations, $c=1,$ and the limit allowing an exact result, $c=c_{ex}.$ We
set $r=-1.2$ and $E=1.5$ for Fig.3. Hence $c_{ex}=-1.29$ and $D_{ex}=0.51.$
The full lines show two branches of estimates provided by $R_{cc}.$
Its other two branches turn out to be either complex or out of range in that
case. Not surprisingly, one of the full lines contains the point
$(c_{ex},D_{ex}),$ representing the fully symmetric, exact solution
$\Psi,\Psi'.$

It is remarkable that the long-dashed line, which corresponds to the roots of
$R_{tb},$ also contains this exact point. It is also remarkable that the
time reversal breaking, measured by $|s-u|$ for instance,
stays finite on that long-dashed branch when $c$ approaches
$c_{ex}$ and does not vanish at that point. It is nevertheless easy to verify
that, when $c=c_{ex},$ then $F_{tb}\equiv D_{ex}$ as a function of $E$ and $r.$
This indicates that, as long as correlations are allowed in the trial
states, even partly only, the breaking of time reversal symmetry may help in
finding good estimates of matrix elements of propagators. The paradox of a
finite value of $s-u$ at the exact point is understandable, because a
well-known property of $F,$ see Eq.(1), is that only any one of the two
equations, Eqs.(2), needs to be solved exactly to obtain $D_{ex}.$ It is
unfortunate that, as shown by Eq.(9), the stationary value $F_{tb}$ provided
by uncorrelated trial states, namely for $c=1,$ identically vanishes.

The second full line branch goes through that same point $(c,D)=(1,0).$
This is no accident, because $R_{cc}$ gets a trivial, energy independent root
$s=-r$ when $c=1,$ for which the fully symmetric functional identically
vanishes. Only three fully symmetric saddle points are left to depend on $E$
for uncorrelated trial states, together with the {\it pb} branch. In the
present case, besides that trivial fourth fully symmetric root $s=-r=1.2,$
one fully symmetric root is real and the other two are complex.

The short-dashed line on Fig.3 represents the behavior of the {\it pb}
branch. It shows a vague trend to interpolate between the two real,
fully symmetric branches, but its utility will be apparent later only. Indeed,
it does not go through the exact point $(c_{ex},D_{ex}),$ nor does it show
an accurate reproduction of $D_{ex}$ for $c=1.$

We show on Fig.4 the multivalued nature of the estimate(s) provided by the
four roots of $R_{cc}.$ Obviously, the number of real roots depends on
$E,r,c.$

We now turn to behaviors of the estimates of $D$ as functions of $E.$ This is
illustrated by Fig.5, where $D_{ex}(E)$ (full line) is compared to
its various estimates with $c=1.$ It is seen at once that a fully symmetric
estimate (dashed line) is convenient to reproduce pole behaviors near
$E=\pm 2.$ This is not completely surprising since the corresponding
eigenstates $|a^2>$ and $|b^2>$ are themselves invariant under particle
exchange. Conversely, eigenstates such as $|a_1b_2>$ and $|b_1a_2>$ are
not invariant under such an exchange, and it is thus reasonable that
a {\it pb} estimate is needed to account for the pole at $E=0.$ Moreover, it
is consistent that the $\it pb$ estimate fails for $E \simeq \pm 2,$ and the
fully symmetric estimate in turn fails for $E \simeq 0.$ For the sake of
completeness, we also show on Fig.5 the behavior of the {\it tb} estimate
obtained when $c=0.$ It is clear that this {\it tb} estimate, like the
{\it pb} one, succeeds for $E\simeq 0$ and fails for $E\simeq \pm 2.$ All these
properties are transparent from the corresponding special values of the
functionals,
\begin{mathletters} \begin{eqnarray}
F_{pb}|_{c=1} &=& \frac{(Er^2+E+2)(2r^2-Er^2-E)}{4 E}, \\
F_{tb}|_{c=0} &=& \frac{2r^2}{E}, \\
F_{cc}|_{c=1} &=& \frac{(r+s)^4}{(s^2+1)(Es^2-2s^2+E+2)}.
\end{eqnarray} \end{mathletters}
The residues of the special values, Eqs.(10a,b), of $F_{pb}$ and $F_{tb}$
at $E=0$ are $r^2$ and $2r^2,$ respectively. This compares well with the
exact residue $2r^2,$ see Eq.(4). The three non trivial saddle points of
$F_{cc}|_{c=1}$ are the roots of
\begin{equation}
R_{cc}|_{c=1}=r(2-E)s^3+Es^2-Ers+(2+E).
\end{equation}
A straightforward elimination of $s$ between Eq.(11) and Eq.(10c)
generates a cubic polynomial condition relating these three saddle point
estimates $D_{cc}$ to $r$ and $E,$
\widetext \begin{equation}
%8 + 12*e + 6*e^2 + e^3 + (24*e + 24*e^2 + 6*e^3)*r^2 +
%  (-24 - 12*e + 30*e^2 + 15*e^3)*r^4 + (-48*e + 20*e^3)*r^6 +
%  (24 - 12*e - 30*e^2 + 15*e^3)*r^8 + (24*e - 24*e^2 + 6*e^3)*r^10 +
%  (-8 + 12*e - 6*e^2 + e^3)*r^12 +
%  (48 + 48*e + 8*e^2 - 4*e^3 - e^4 + (-48*e - 24*e^2 - 8*e^3 - 4*e^4)*r^2 +
%     (336 - 64*e^2 - 6*e^4)*r^4 + (48*e - 24*e^2 + 8*e^3 - 4*e^4)*r^6 +
%     (48 - 48*e + 8*e^2 + 4*e^3 - e^4)*r^8)*z +
%  (96 + 48*e - 16*e^2 - 8*e^3 + (-192*e + 48*e^3)*r^2 +
%     (-96 + 48*e + 16*e^2 - 8*e^3)*r^4)*z^2 + (64 - 16*e^2)*z^3
%8 + 12 E + 6 E^2 + E^3 + (24 E + 24 E^2 + 6 E^3) r^2 +
%  (-24 - 12 E + 30 E^2 + 15 E^3) r^4 + (-48 E + 20 E^3) r^6 +
%  (24 - 12 E - 30 E^2 + 15 E^3) r^8 + (24 E - 24 E^2 + 6 E^3) r^10 +
%  [-8 + 12 E - 6 E^2 + E^3) r^12 +
%  (48 + 48 E + 8 E^2 - 4 E^3 - E^4 + (-48 E - 24 E^2 - 8 E^3 - 4 E^4) r^2 +
%     (336 - 64 E^2 - 6 E^4) r^4 + (48 E - 24 E^2 + 8 E^3 - 4 E^4) r^6 +
%     (48 - 48 E + 8 E^2 + 4 E^3 - E^4) r^8] z +
16(4-E^2)D_{cc}^3+8[12+6E-2E^2-E^3+6(E^2-4)Er^2+(-12+6E+2E^2-E^3)r^4]D_{cc}^2
+\alpha D_{cc} + \beta = 0,
\end{equation} \narrowtext \noindent
where $\alpha$ and $\beta$ depend on $r$ and $E,$ naturally. For $E^2=4$ this
polynomial, Eq.(12), reduces to degree 2 instead of 3, and the multivalued
$D_{cc}$ has indeed first order poles for $E=2$ and $E=-2,$ with residues
$1$ and $r^4,$ respectively. These agree with the exact residues, see Eq.(4).

Finally on Fig.6 we compare, for r=1.1, the exact $D,$ as a function of $E,$
with two kinds of estimates obtained with $c=1.$ The small-dashed line
indicates the real root of Eq.(12), already found to show poles
at $E=\pm 2.$ The long-dashed line shows the real part of the other two roots
of Eq.(12). Its complete lack of pole structure makes it much less valuable
than the real root.

In conclusion, this letter gives three results for the variational theory
of propagators. The first one is the existence of a non trivial {\it and}
soluble model for mean field approximations, which are notoriously non linear
and demand a sorting of their (intricate) solutions. The second result is
indeed a complete classification of the solution branches of the model,
with their physical meaning, namely symmetry conservation or breaking.
We can keep track of their interconnection or lack of connection. We
can also assess the validity domain of the amplitude estimate given by each
branch. Of special interest are the time reversal breaking solutions which we
discovered, because they can provide the exact amplitude and exist only if
enough correlation is included in the trial states. When correlations are
excluded, parity breaking solutions still exist and are absolutely necessary
to account for several singularities of the exact amplitude. Fully
symmetric solutions are not flexible enough, in the uncorrelated variational
space, to account for all the singularities of the exact, fully symmetric
solution, which is unique and correlated. The third result is then the
necessity of a linear admixture of such solution branches, to remove any
ambiguity of choices between them, patch their limited validity domains,  and
restore the uniqueness of the physical solution.
As a first step for the representation of physical correlations, mean field
solutions can thus be better analyzed along the lines described by this model.

{\it Acknowledgements:} A.W. thanks Service de Physique Th\'eorique
for its hospitality during part of the work.

\begin{figure}[htb] \begin{center}
\mbox{ \epsfysize=65mm
%%%arXiv         \epsffile{Fig1.eps}
     }
\caption{Contour plot of $F,$ see Eq.(6), for $(s,t)=(u,v)$ when
$(r,E,c)=(1,3,-0.25).$ Vicinity of parity breaking saddle point
$(s,t)=(-1.19,0.13)$ with saddle value $F_{pb}=0.46.$}
\end{center} \end{figure}
\begin{figure}[htb] \begin{center}
\mbox{ \epsfysize=65mm
%%%arXiv        \epsffile{Fig2.eps}
     }
\caption{Contour plot of $F$ for $(r,E,c)=(1,1,-0.25)$ and
$(s,u)=(t,v).$ Vicinity of the time reversal breaking saddle point
$(s,u)=(-0.73,-0.068)$ with saddle value $F_{tb}=1.58.$}
\end{center} \end{figure}
\begin{figure}[htb] \centering
\mbox{  \epsfysize=100mm
%%%arXiv         \epsffile{Fig3.eps}
     }
\caption{Various estimates of amplitude $D$ as functions of the constraint $c$
for $(r,E)=(-1.2,1.5).$ Full lines: full symmetry. Small-dashed line: parity
breaking. Long-dashed line: time reversal breaking.
Exact $D_{ex}=0.51$ at $c_{ex}=-1.29.$}
\end{figure}

\begin{figure}[htb] \centering
\mbox{ \epsfysize=80mm
%%%arXiv        \epsffile{Fig4.eps}
     }
\caption{Fully symmetric estimates as functions of $c$ for $(r,E)=(1,1.5).$
Any folding of one of the full lines onto itself means a transition between
two and four real estimates.}
\end{figure}

\begin{figure}[htb] \centering
\mbox{ \epsfysize=100mm
%%%arXiv        \epsffile{Fig5.eps}
     }
\caption{Amplitudes as functions of $E$ for $r=-1.2.$ Full line: $D_{ex}.$
Dashed line: full symmetry estimate, $c=1.$ Short-dashed: {\it pb,} $c=1.$
Long-dashed: {\it tb,} $c=0.$ Poles demand full symmetry near $E=\pm 2$ and
{\it pb} or {\it tb} near $E=0.$}
\end{figure}

\begin{figure}[htb] \centering
\mbox{ \epsfysize=100mm
%%%arXiv        \epsffile{Fig6.eps}
     }
\caption{Full symmetry estimates as functions of $E$ for $(r,c)=(1.1,1).$ Short
dashes: $F$ real. Long dashes: $ReF,$ complex $F.$ Full line: $D_{ex}.$
Poles $E=\pm 2$ recovered by $F$ real.}
%\label{Klineausr}
\end{figure}

\end{document}